\documentclass[12pt,preprint]{aastex}
\usepackage{natbib}
\usepackage{graphicx}
\newcommand{\E}{\times 10^}
\received{2004 August 5}
\begin{document}

\title{Ultraviolet HST Observations of the Jet in M87\altaffilmark{1}}
\author{Christopher Z. Waters and Stephen E. Zepf}
\affil{Department of Physics and Astronomy, Michigan State University,
  East Lansing, MI 48824}
\email{watersc1@pa.msu.edu,zepf@pa.msu.edu}

\altaffiltext{1}{Based on observations made with the NASA/ESA Hubble
  Space Telescope, obtained at the Space Telescope Science Institute,
  which is operated by the Association of Universities for Research in
  Astronomy, Inc., under NASA contract NAS 5-26555.}

\begin{abstract}
  We present new ultraviolet photometry of the jet in M87 obtained
  from HST WFPC2 imaging.  We combine these ultraviolet data with
  previously published photometry for the knots of the jet in radio,
  optical, and X-ray, and fit three theoretical synchrotron models to
  the full data set.  The synchrotron models consistently overpredict
  the flux in the ultraviolet when fit over the entire dataset.  We
  show that if the fit is restricted to the radio through ultraviolet
  data, the synchrotron models can provide a good match to the data.
  The break frequencies of these fits are much lower than previous
  estimates.  The implied synchrotron lifetimes for the bulk of the
  emitting population are longer than earlier work, but still much
  shorter than the estimated kinematic lifetimes of the knots.  The
  observed X-ray flux cannot be successfully explained by the simple
  synchrotron models that fit the ultraviolet and optical fluxes.  We
  discuss the possible implications of these results for the physical
  properties of the M87 jet.  We also observe increased flux for the
  HST-1 knot that is consistent with previous results for flaring.
  This observation fills in a significant gap in the time coverage
  early in the history of the flare, and therefore sets constraints on
  the initial brightening of the flare.
\end{abstract}


\section{Introduction}

M87 is one of the closest galaxies hosting a prominent jet.  Its
location in the core of the Virgo cluster at a distance of 16 Mpc
(e.g. \citealt{Macri}) makes M87 close enough that the structure of the
jet can be resolved by space-based observations.  It has been studied
extensively from the radio (e.g., \citealt{SBM}, \citealt{BZO}) to X-ray
(e.g., \citealt{Marshall}, \citealt{Wilson}), with many broadband studies
in the optical (e.g. \citealt{Perlman}, \citealt{BSH}).  In addition to
photometric surveys of the jet, polarimetry maps of the jet have been
created \citep{Perlman_polarimetry}, and the bulk motion of the jet
has been investigated by measuring the proper motions of the jet's
component knots over decade-long observation programs \citep{BZO}.
All of these observations have provided more information about M87
than is known for nearly any other extragalactic jet.  

However, there have been no far ultraviolet observations with spatial
resolution sufficient to resolve the structure of the jet in detail.
These observations are essential for constraining the synchrotron
emission, because the break frequencies of the models fit to extant
data are between the optical and X-ray.  Moreover, the different
models that fit the radio, optical, and X-ray data make significantly
different predictions for the shape of the spectrum in the ultraviolet.

In this paper, we use high spatial resolution ultraviolet images
obtained with HST to address questions about the physical
characteristics of the emission from the M87 jet.  In Section 2, we
present these observations and detail the reduction process.  We
compare our measurements to theoretical models of the jet emission in
Section 3.  In Section 4 we discuss the evidence for flaring in our
observations of HST-1, and use this to constrain the early evolution
of this flare.  We summarize our conclusions in section 5.

\section{Observations and Analysis}

The jet of M87 was observed with the WFPC2 detector aboard HST on
2001 February 23.  The galaxy core was centered in the PC, with the
jet oriented along the columns of the detector.  The observations were
obtained through the F170W filter, which has an effective wavelength
of $1666$\AA{}.  Six images were taken in an ``L'' shaped dither
pattern with $0\farcs25$ shifts between the positions.  The total
integration time of the observations was 7600 seconds.

We combined the six observations with the Dither package in IRAF,
using the DRIZZLE algorithm from \citet{fruchterandhook}.  This both
cleans the final image of cosmic rays and CCD flaws and takes
advantage of the subpixel spacing between image positions to construct
a final image with smaller pixels.  The Dither package also corrects
the input images for the geometric distortion of the WFPC2, which is
largest in the ultraviolet.

The knots of the jet were photometered based on the apertures defined
by \citet{Perlman} in their study of HST optical images of the jets.
We have chosen to use the same apertures to allow direct comparison of
our fluxes with those presented.  The boundaries of our apertures are
defined by finding the closest pixel boundary that matches the
positions listed by Perlman.  In order to ensure that the core of the
galaxy does not interfere with any of the knots, we have removed the
galaxy profile using the ISOPHOTE package in IRAF.  Only knot HST-1
has any possible contribution from the core, as there is little
evidence of the galaxy beyond $0\farcs55$ from the core.

The photometry determined within these apertures was calibrated using
the default values for the F170W filter given in the WFPC2 Calibration
Manual \citep{WFPC2}.  Because our images have low background levels,
we need to account for possible charge transfer efficiency losses.  We
calculated the correction using methods from both \citet{WHC} and
\citet{Dolphin}.  As the results were similar, we used the average of
the two methods as our final correction, and the difference between
the methods as the uncertainty for this correction.  Since both of
these methods are based on the assumption of a point source object, we
adopted the suggestion presented in \citet{Riess} to half the
calculated correction.  The typical level of this correction was about
$10\%$.  Another possible concern is the effect of contaminants on the
WFPC2 is generally worse in the UV than at other wavelengths.  For
this set of observations, the WFPC2 had been decontaminated less than
a week prior, so the correction we applied is less than $2\%$ of the
measured counts.  We used no aperture correction for these
measurements, as our apertures extend well beyond the visible portions
of the knots.  The final fluxes, along with the distance from the core
for each knot are listed in Table \ref{tab:data}.

\clearpage
\begin{deluxetable}{ccc}
\tabletypesize{\small}
\tablewidth{0pc}
\tablecaption{Photometric Data\label{tab:data}}
\tablehead{\colhead{Knot} & \colhead{F(F170W) ($\mu$Jy)} & Distance (arcsec)}
\startdata
HST-1 & $15.8 \pm 0.6$ &  1.25\\
D     & $26.7 \pm 0.9$ &  3.45\\
D-East& $17.8 \pm 1.1$ &  2.85\\
D-Mid & $ 4.7 \pm 0.4$ &  3.52\\
D-West& $ 5.7 \pm 0.5$ &  3.52\\
E     & $10.0 \pm 0.9$ &  6.30\\
F     & $34.5 \pm 1.7$ &  8.72\\
I     & $11.3 \pm 1.4$ & 11.19\\
A     &$275.2 \pm12.7$ & 12.67\\
A-bar & $15.3 \pm 2.9$ & 11.80\\
A-shock&$166.2\pm 9.6$ & 12.40\\
A-diff& $86.9 \pm 6.7$ & 13.28\\
B     &$147.4 \pm 7.9$ & 14.98\\
B-1   & $83.9 \pm 6.5$ & 14.41\\
B-2   & $44.3 \pm 3.9$ & 15.59\\
C-1   & $80.8 \pm12.9$ & 17.53
\enddata
\end{deluxetable}
\clearpage
\section{Comparison to Models}

We combined our ultraviolet photometry with published radio and
optical data from \citet{Perlman} and X-ray data from
\citet{Marshall}.  We chose three standard theoretical synchrotron
models to fit to the data.  The KP model \citep{Kardashev,Pacholczyk}
assumes that the source of the emission is a single burst of energetic
electrons with an isotropic pitch angle distribution.  The pitch angle
distribution of this model is not allowed to vary, so some high energy
electrons can remain with trajectories nearly parallel to the magnetic
field.  Although this model is physically unlikely, as relativisitic
electrons are known to scatter, we include this model in our fits to
compare with previously published results using this model.  The JP
model \citep{jp} also assumes a single burst of electrons with an
isotropic pitch angle distribution, but allows the pitch angle
distribution to scatter so that it always has an isotropic
distribution.  This allows all high energy electrons to quickly
radiate away their energy, resulting in an exponential falloff beyond
the break frequency.  We also consider a continuous injection (CI)
model of \citet{CI} in which energetic electrons are being constantly
added to the emitting region.  Although other synchrotron models have
been considered for M87 \citep[e.g.][]{CB} , these models are the most
common for comparison to the data, and are likely to bracket the range
of physical possibilities.

\begin{figure}
  \figurenum{1}
  \plotone{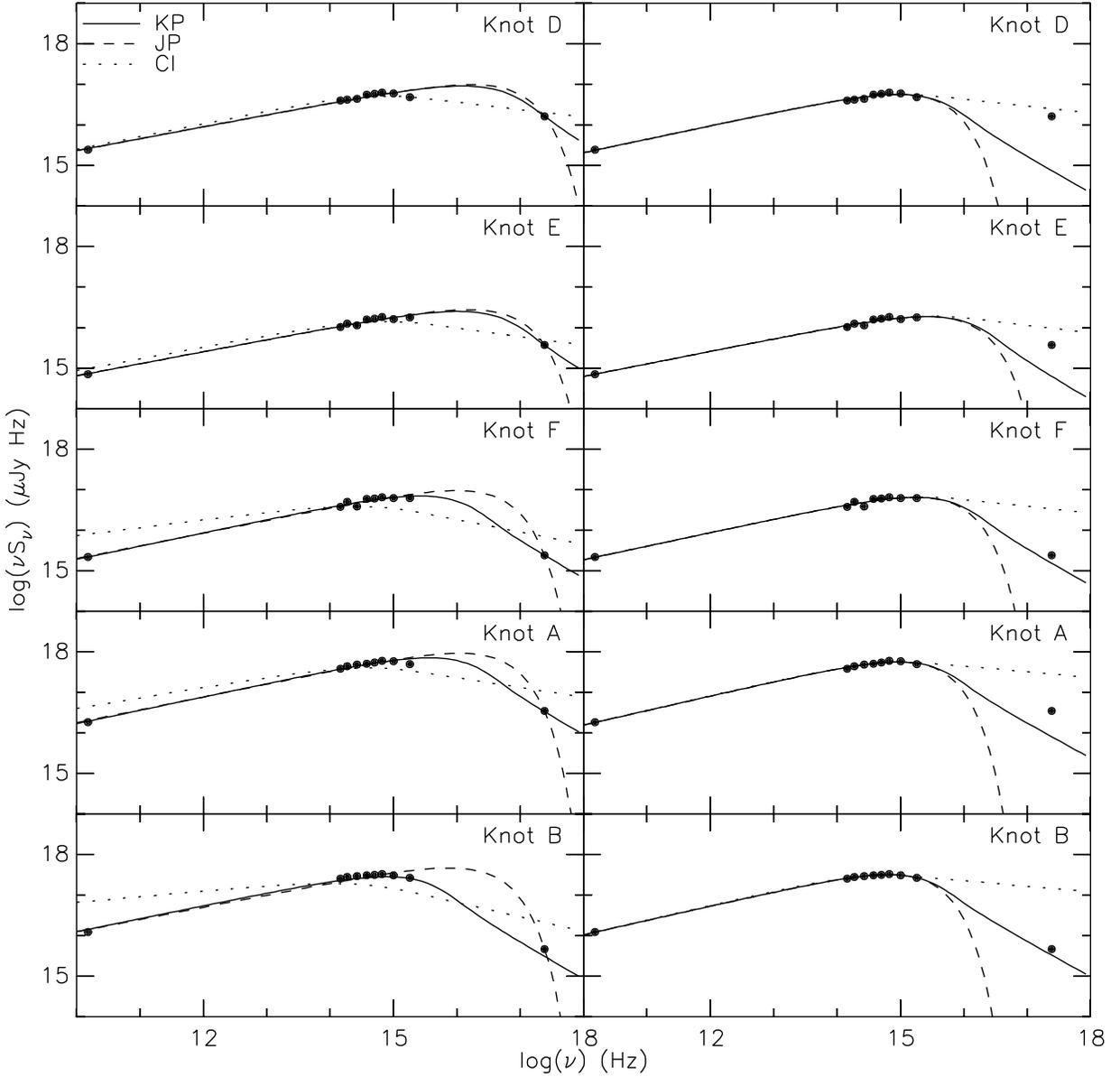}
  \caption{A plot of the spectral energy distributions for the knots
    of the jet from radio through X-ray wavelengths.  The data are
    plotted as points, and the best fits for each of the synchrotron
    models are plotted as shown above.  The error bars for the fluxes
    are smaller than the points used.  The left panel displays model
    fits that include the X-ray data, and the right panel shows the
    fits without these data.  The left panel demonstrates that the
    models are unable to fit both the turnover from the power law seen
    in the ultraviolet and the X-ray data point.  The right panel
    shows that the synchrotron models with much lower break
    frequencies can provide a reasonable fit to the radio through
    ultraviolet data.  The best fitting parameters of these models are
    listed in Table \ref{tab:model} and Table \ref{tab:modelX}}
  \label{fig:All}
\end{figure}

The synchrotron fitting to the data was performed using new code based
on a program provided by Chris Carilli \citep{Carilli}.  Our code
determines the best fitting low frequency power law index, break
frequency, and intensity scaling to the data using the
Levenberg-Marquardt algorithm implementation in the GNU Scientific
Library\footnote{http://www.gnu.org/software/gsl/}.  Using this
procedure, we determine the best fit parameters for each model over
the full spectral range.  These best fit values, along with the
reduced chi-squared values for each knot fit are listed in Table
\ref{tab:model}.  For reference, the kinematic ages calculated from
the proper motions \citep{BZO} of the knots are listed where
available.  The final calculated best fitting model SEDs are shown in
the left panels of Figure \ref{fig:All}.

The models described above appear to be unable to fit the full
optical, ultraviolet and X-ray spectrum for many of the knots.  They
systematically overpredict our far UV flux, and to a lesser extent,
the previously published near ultraviolet (F300W) data.  The flux in
the ultraviolet also appears to drop below the simple power-law form
produced by synchrotron models at frequencies well below the break
frequency.  This ultraviolet turnover suggests that we are close to
the break frequency for a synchrotron model that fits the radio to
ultraviolet data.  If this is the case, then the X-ray flux is higher
than the prediction for a simple synchrotron model, and it would seem
that this fact is forcing our code to choose higher break frequency
models, despite their poor fit in the ultraviolet.  By refitting the
models without the X-ray flux, we can generate models that are much
better fits to the radio through ultraviolet emission.  The best
fitting parameters for these new fits are given in table
\ref{tab:modelX}, and the new synchrotron fits are shown in the right
panels of Figure \ref{fig:All}.  These fits provide much
lower break frequencies for the emission, and for the three knots with
obvious ultraviolet turnovers (knots D, A, and B), these fits have
much lower $\chi^2$ values compared to those generated using the X-ray
measurement.  The remaining two knots (knots E and F), have minimal
difference in the fit quality, which suggests that we still do not
have direct measurements around the break frequency.

If the radio through far UV emission is generated by a single burst
synchrotron model, then the excess of X-ray flux compared to these
models must be accounted for.  Inverse Compton is a tempting option,
as it has been used to explain X-ray emission in other jets.  However,
previous analysis has shown that the geometry of the jet required for
inverse Compton to contribute significantly to the X-ray flux from the
knots is inconsistent with the geometry indicated by proper motion
studies.  Specifically, the beaming model presented by \citet{HK}
predicts that in order to explain the observed X-ray flux, the angle
of the jet to the line of sight must be nearly zero, in contradiction
to the $20^\circ$ value derived from proper motions \citep{BSM}.  It
is clear then that there must be some reinjection or re-excitation of
electrons in the emitting regions.  This scenario has some direct
evidence to support it.  The kinematic ages of the knots in the jet
are much longer than the calculated synchrotron ages.  Therefore, even
to explain the optical emission of the jet, some amount of fresh high
energy material from shocks along the jet must be added. 
\clearpage
\begin{deluxetable}{lccccccc}
\tabletypesize{\small}
\tablewidth{0pc}
\tablecaption{Parameters for Synchrotron Model Fits for Radio through X-ray Data\label{tab:model}}
\tablehead{\colhead{} &\colhead{} & \colhead{HST-1} & \colhead{D} &
  \colhead{E} & \colhead{F} & \colhead{A} & \colhead{B}}
\startdata
    & $T_{kin}$(yr) &  & 2037 &  & 2500 & 6169 & 6353 \\
    \hline
 KP:& $\alpha$ & $0.711$ & $0.706$ & $0.703$ & $0.671$ & $0.670$ & $0.651$ \\
    & $\nu_{B}$(Hz)& $7.52\E{15}$ & $5.15\E{16}$ & $4.32\E{16}$ 
    & $8.80\E{15}$ & $1.22\E{16}$ & $2.71\E{15}$ \\
    & $\chi^2$& $3.22\E{3}$ & $39.1$ & $8.91$ & $11.1$ & $14.4$ & $3.21$ \\
    & $\tau_{sync}$(yr)& $386$ & $148$ & $161$ & $357$ & $303$ & $643$ \\
    \hline
 JP:& $\alpha$ & $0.718$ & $0.708$ & $0.706$ & $0.686$ & $0.683$ & $0.699$ \\
    & $\nu_{B}$(Hz)& $1.00\E{16}$ & $1.24\E{17}$ & $1.12\E{17}$ 
    & $6.35\E{16}$ & $6.99\E{16}$ & $5.21\E{16}$ \\
    & $\chi^2$& $1.29\E{4}$ & $42.1$ & $9.48$ & $15.4$ & $32.4$ & $54.2$ \\
    & $\tau_{sync}$(yr)& $335$ &  $95$ & $100$ & $134$ & $127$ & $147$ \\
    \hline
 CI:& $\alpha$ & $0.705$ & $0.683$ & $0.711$ & $0.804$ & $0.741$ & $0.855$ \\
    & $\nu_{B}$(Hz)& $1.56\E{21}$ & $9.27\E{14}$ & $1.22\E{15}$ 
    & $8.75\E{14}$ & $7.60\E{14}$ & $6.48\E{14}$ \\
    & $\chi^2$& $37.2$ & $44.3$ & $83.6$ & $5.62\E{3}$ & $2.86\E{3}$ & $2.65\E{4}$ \\
    & $\tau_{sync}$(yr)& $1$ &  $1100$ & $959$ & $1132$ & $1214$ &
    $1315$
\enddata
\end{deluxetable}
\begin{deluxetable}{lcccccc}
\tabletypesize{\small}
\tablewidth{0pc}
\tablecaption{Parameters for Synchrotron Model Fits for Radio through Ultraviolet Data \label{tab:modelX}}
\tablehead{\colhead{} &\colhead{} & \colhead{D} &   \colhead{E} &
  \colhead{F} & \colhead{A} & \colhead{B}}
\startdata
 KP:& $\alpha$ & $0.668$ & $0.690$ & $0.664$ & $0.641$ & $0.645$ \\
    & $\nu_{B}$(Hz)& $3.07\E{15}$ & $9.77\E{15}$ 
    & $5.61\E{15}$ & $2.66\E{15}$ & $1.90\E{15}$ \\
    & $\chi^2$& $9.81$ & $8.23$ & $12.7$ & $3.15$ & $0.54$ \\
    & $\tau_{sync}$(yr)& $604$ & $339$ & $447$ & $649$ & $768$ \\
    \hline
 JP:& $\alpha$ & $0.668$ & $0.690$ & $0.664$ & $0.642$ & $0.646$ \\
    & $\nu_{B}$(Hz)& $5.36\E{15}$ & $1.68\E{16}$ 
    & $9.70\E{15}$ & $4.70\E{15}$ & $3.43\E{15}$ \\
    & $\chi^2$& $9.63$ & $8.22$ & $12.7$ & $3.04$ & $0.50$ \\
    & $\tau_{sync}$(yr)&  $457$ & $258$ & $340$ & $488$ & $572$ \\
    \hline
 CI:& $\alpha$ & $0.668$ & $0.690$ & $0.664$ & $0.639$ & $0.636$ \\
    & $\nu_{B}$(Hz)& $1.19\E{15}$ & $4.22\E{15}$ 
    & $2.32\E{15}$ & $9.43\E{14}$ & $5.10\E{14}$ \\
    & $\chi^2$& $11.9$ & $8.35$ & $13$ & $4.27$ & $1.82$ \\
    & $\tau_{sync}$(yr)&  $ 971$ & $515$ & $ 695$ & $1090$ & $1483$
\enddata
\end{deluxetable}
\clearpage

\clearpage
\section{Flaring Outburst of HST-1}

\begin{figure}
  \figurenum{2}
  \plotone{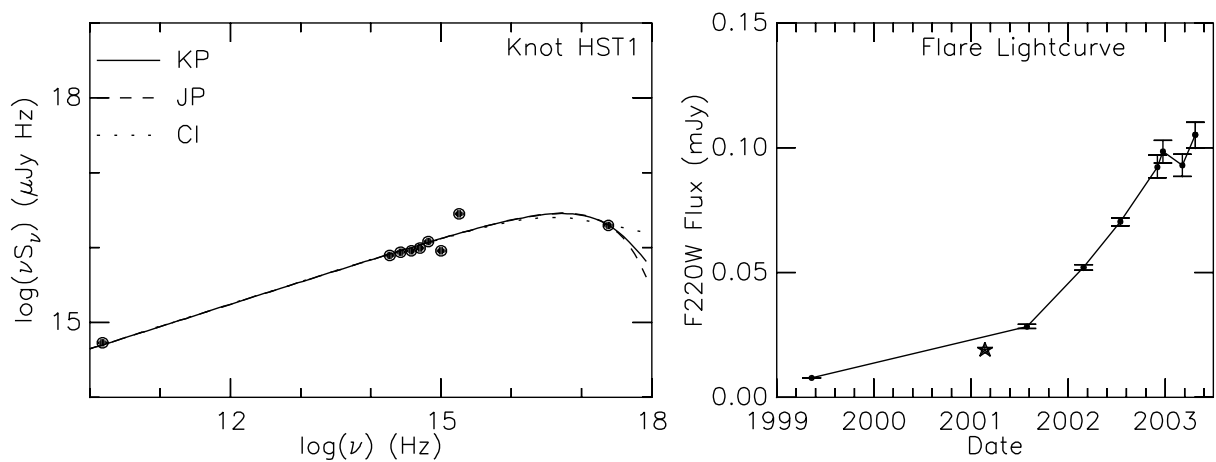}
  \caption{Best fitting synchrotron SED for HST-1.  The new UV point
    can be seen to exceed the fit by a large amount.  The right panel
    shows the light curve for the knot HST-1 based on data from
    Perlman and Biretta (private communication).  The new data point
    is plotted as a star.}
  \label{fig:flare}
\end{figure}

HST-1, the knot closest to the core of M87, has been shown to have a
very dynamic light curve \citep{Perlman_flare,Harris_flare}.
Observations from late 2001 to the present show a significant increase
in the brightness of HST-1 compared to initial optical observations on
1999 May 11 and 1998 February 25 \citep{Perlman,Perlman_flare}, along
with shorter timescale variability.  As can be seen in the right panel
of Figure \ref{fig:flare}, our observation of HST-1 obtained in
February 2001 fills a significant gap in the light curve of the flare,
and thus it is interesting to examine our data to help constrain the
early evolution of the flare.

By comparing our ultraviolet photometry for HST-1 to the synchrotron
model fits of earlier optical coeval data \citep{Perlman} we can
immediately note that our observation is substantially brighter than
expected from these earlier data, as shown in the left panel of
Figure \ref{fig:flare}.  As the fluxes for the other knots in the jet
tend to either lie in agreement with the synchrotron fits, or suggest
a drop in the ultraviolet, it is unlikely that this brightening is a
result of an error in the data reduction.  This suggests that our
observation shows that HST-1 has begun its flaring outburst by the
time of our observations.

To make a quantitative comparison to the many observations of knot
HST-1 obtained at different times, we need to convert all measurements
to the flux at a standard wavelength.  We follow previous work on this
flare \citep{Perlman_flare} and adopt a power law index of $\alpha =
0.6$ to scale our measurement to a standard wavelength of $2200$\AA{}.
Applied to our data through the F170W filter gives a flux of
$F_{220nm} = 18.67 \pm 0.71 \mu$Jy in our measurement on 2001 February
23.  A plot of this measurement and those of the flare of HST-1 at
other times are shown in figure \ref{fig:flare}b.  Our data requires
the flare to have begun before February 2001, significantly pushing
back the earliest observation of the flare.  Adopting a simple
exponential function to fit the initial rise of the flare gives a flux
doubling time scale for the beginning of the flare of 283 days,
consistent with other previously published timescales, and constrains
the start of the flare to February 2000.

The flaring emission in HST-1 arises from the upstream edge of the
knot.  This region is consistent with a source size below our
resolution limit, giving us a maximum size for the flaring region of
1.8pc.  If we assume that our flux doubling timescale is on the order
of the light crossing time, then we can set an upper limit on the
Doppler factor for the emission region of $\delta \lesssim 8$, as a
value any larger would allow us to resolve the source of the flare.
This result is consistent with multiwavelength results presented in
\citet{Perlman_flare}, where a more detailed analysis has been done of
the constraints the variability places on the physical structure of
the inner jet.  However, as their results are also resolution limited,
it seems clear that higher resolution imaging is required to further
constrain the size of the flaring outburst.

\clearpage
\clearpage

\section{Discussion}
The standard synchrotron models that are used to fit extragalactic
jets all predict a power law from the radio through the optical, and
have a break in the emission somewhere between ultraviolet and X-ray
frequencies.  These models can readily fit both the optical power law
and an X-ray measurement by simply choosing a sufficiently high break
frequency.  The ultraviolet flux predicted by these fits is high, as
the power law extends into the ultraviolet for such high break
frequencies.  However, our ultraviolet observations of the knots of
the M87 jet have shown that we have detected a turnover in the
spectrum for knots D, A, and B.  The break frequencies that
accurately model these turnovers are much lower than those previously
determined to fit the X-ray flux.  This leads to a problem in using
these existing synchrotron models to fit the full radio to X-ray
spectrum.  

The single burst models we have considered significantly underpredict
the X-ray flux with the lower break frequencies required by the
ultraviolet data.  The CI model has the opposite problem, as it
overpredicts the X-ray flux.  For the majority of the knots, the
spectral break is much larger than that predicted by the CI model,
consistent with the steep X-ray spectra found by \citet{Wilson}.  Only
knot D has a spectral break close to the CI model prediction of
$\Delta\alpha = 0.5$.  The X-ray data from \citet{Wilson} also
indicates that the emission peak for the X-ray data is further
upstream than the optical and radio emission peaks.  This observation
supports a scenario in which the electrons are shocked at the upstream
edge of the knot, and then spectrally age as they travel downstream.
Therefore, the fact that the CI model does not accurately predict the
high frequency behavior may suggest that the properties in the
emitting region are not as static as the CI model assumes.  Changes in
the downstream properties like those studied by \citet{CB} can give
rise to large breaks similar to what we measure.

Another possible explanation for the observed spectra arises from
allowing the number of electrons excited to vary over time.  The JP
model with its single burst, and the continuous injection of the CI
model can be considered limiting cases of such a variation.  As the
observed spectra are intermediate between these two models, but lie
closer to that predicted by the CI model, it is possible that the
shock fronts simply do not always excite the same number of electrons.
The flaring emission of knot HST-1 indicates that the emission from
the shock is not necessarily constant in time, and that variation in
the shock as well as downstream may contribute to the deviations from
the theoretical model.  Our observation of this knot helps to
constrain the early evolution of the flare, suggesting a start date
for the flare in early February 2000, and giving a doubling time scale
of 0.77 years, similar to that found at later times for the flare.
This flaring behavior illustrates that the emission of the knots in
the jet is influenced by shocks on scales smaller than the observed
sizes of the knots.

We thank E. Perlman for helpful discussions and for supplying data for
the light curve of knot HST-1 and C. Carilli for making his
synchrotron fitting code available.  We also thank the anonymous
referee for helpful comments and discussion.  This work was supported
by NASA through grant HST-GO-08725-05-A and awarded by the Space
Telescope Science Institute, which is operated by the Association of
Universities for Research in Astronomy, Inc., for NASA under contract
NAS5-26555.

\end{document}